# Numerical investigation of MILD combustion using Multi-Environment Eulerian Probability Density Function modeling

**Author(s)**
Akshay Dongre[1], Ashoke De[1], Rakesh Yadav[2]

[1]Department of Aerospace Engineering, Indian Institute of Technology, Kanpur, India-208016

[2]Ansys Fluent India Pvt. Ltd., Pune, India-411057

**Abstract**
In the present paper, the flames imitating Moderate and Intense Low Oxygen Dilution (MILD) combustion are studied using the Probability Density Function (PDF) modeling approach. Two burners which imitate MILD combustion are considered for the current study: one is Adelaide Jet-in-Hot-Coflow (JHC) burner and the other one is Delft-Jet-In-Hot-Coflow (DJHC) burner. 2D RANS simulations have been carried out using Multi-environment Eulerian Probability Density Function (MEPDF) approach along with the Interaction-by-Exchange-with-Mean (IEM) micro-mixing model. A quantitative comparison is made to assess the accuracy and predictive capability of the MEPDF model in the MILD combustion regime. The computations are performed for two different jet speeds corresponding to Reynolds numbers of Re=4100 and Re=8800 for DJHC burner, while Re=10000 is considered for the Adelaide burner. In the case of DJHC burner, for Re=4100, it has been observed that the mean axial velocity profiles and the turbulent kinetic energy profiles are in good agreement with the experimental database while the temperature profiles are slightly over-predicted in the downstream region. For the higher Reynolds number case (Re=8800), the accuracy of the predictions is found to be reduced. Whereas in the case of Adelaide burner, the computed profiles of temperature and the mass fraction of major species ($CH_4$, $H_2$, $N_2$, $O_2$) are found to be in excellent agreement with the measurements while the discrepancies are observed in the mass fraction profiles of $CO_2$ and $H_2O$. In addition, the effects of differential diffusion are observed due to the presence of $H_2$ in the fuel mixture.

## 1. INTRODUCTION

Moderate and Intense Low Oxygen Dilution (MILD) combustion is well known as one of the clean combustion techniques because of its inherent advantages. In particular, this method has evolved as one of the promising techniques for highly energy efficient, low-noise, and low-$NO_x$ industrial combustors. In MILD combustion, the operating conditions are obtained by re-circulating the hot flue gases into the fresh gasses to maintain a very high inlet temperature of the reactants (higher than the auto-ignition temperature), thereby reducing the maximum achievable temperature during the combustion. Therefore the whole combustion process is benefitted in two ways. First, it improves the thermal efficiency of the system through this heat recovery process and secondly it dilutes the oxidizer due to mixing with the hot gases. Due to the reaction with the diluted mixture, the peak temperature remains in the range of 1100-1500 K and hence lowers the $NO_x$ emissions. But, a consequence of these diluted operating conditions is the slow reaction rate that enhances the influence of molecular diffusion on the flame characteristics. This effect, in particular, tests the applicability of the combustion models to simulate the flames in MILD combustion regime and therefore is one of the principle focus of the present study. The various numerical challenges involved in modeling these flames includes the ability to simulate the effects of high re-circulation ratios to ensure proper mixing between reactant stream and flue gases, as well as imitating the recovery of the exhaust gas heat, in other words, imitating the MILD combustion. These are the reasons why this type of combustion has received both experimental [1-4] as well as numerical [5-16] attention.

   Two burners, i.e. Adelaide JHC and DJHC, studied herein fall under the category of MILD combustion. In case of the Adelaide JHC burner, Dally et al. [1] used single-point-Raman-Rayleigh-laser-induced-fluorescence technique to obtain temporally and spatially resolved measurements of temperature and concentration of major species like $CH_4$, $H_2$, $H_2O$, $CO_2$, $N_2$, and $O_2$ and minor species like NO, CO, OH. On the other hand, in case of the DJHC burner, Oldenhof et al. [2-4] used the LDA system and OH-PLIF imaging system to obtain datasets of velocity, temperature, and qualitative OH data on Dutch natural gas flames. The design of the DJHC burner was based on that of the Adelaide JHC burner with main difference being in the design of the secondary burner with Adelaide JHC using $N_2$ to cool down the coflow whereas the DJHC burner uses radiative and convective heat losses along the burner pipe to cool down the coflow. The Adelaide JHC burner experiment was the first of this kind and the DJHC burner was a derivative of that experiment with different measuring techniques and databases which are used in the present study as boundary conditions.

   In the context of Adelaide burner, Christo et al. [5] modeled the methane-hydrogen ($CH_4$-$H_2$) flames using different k-ε turbulence models (Standard k-ε model, Realizable k-ε model, and Renormalization group k-ε model) along with the EDC combustion model with detailed chemistry. The prime focus of their study was to assess the performance of turbulence and combustion model along with various chemical mechanisms. They reported that the standard k-ε (SKE) model with modified dissipation equation constant ($C_{\varepsilon1}=1.6$) produced the best agreement with the experiments, while the differential diffusion effects were found to have strong impact on the accuracy of the predictions for these flames. Despite the reasonable predictions of the flow field, the flame lift-off height was over predicted for the 3% $O_2$ (HM1 flames) case. In another study, Christo et al. [6] modeled these JHC flames using the RANS/PDF hybrid approach. They used Smooke [17], ARM [18], and GRI 3.0 [19] chemical mechanism to describe chemistry along with Euclidean Minimum Spanning Tree (EMST) [20] micro-mixing model to account for molecular mixing. They reported that the JHC flames were kinetically controlled and, hence, the selection of appropriate chemical mechanism will have a major impact on the accuracy of the predictions. It was also emphasized in their work that the use of sophisticated high-temperature/high-oxygen optimized chemical mechanisms need not always improve the accuracy of the predictions but may increase the sensitivity of the solution to slight variations in flow conditions. Kim et al. [7] used the Conditional Moment Closure (CMC) in their study of JHC flames in order to predict the flame structure and NO formation in the MILD regime. Several other RANS based modeling studies conducted on these flames, which include the studies by Frassoldati et al. [8], Mardani et al. [9, 10] and Aminian et al. [11]. In all of these studies, they used EDC combustion model along with DRM 22 [21], GRI 2.11 [22] and KEE-58 [23] chemical mechanisms while the primary focus was to understand the flame structure as well as the importance of molecular diffusion against turbulent transport in the MILD regime. In particular, Aminian et al. [11] obtained substantial improvements in the predictions by increasing the time scale constant in EDC model. This happens because of more distributed reaction zone in MILD regime and the residence time in the fine structures is larger than that in conventional combustion. More recently, Ihme et al. [12, 13] studied these flames using LES methodology while considering the burner as a three-stream problem. They used a Flamelet/Progress Variable (FPV) formulation to introduce an additional conserved scalar in order to efficiently capture the third stream. They primarily emphasized that a third stream is required as a single-mixture fraction was unable to represent the complete coflow mixture composition.

   In context of the DJHC burner, De et al. [14] studied the DJHC flames and reported the effects of different turbulence models in the context of Eddy Dissipation Concept (EDC) turbulence-chemistry interaction model. The major finding of their study is that the chemical time scale needs to be properly controlled to capture the flame behavior in MILD regime. Later on, Kulkarni et al. [15] used Large Eddy Simulation (LES) methodology to study the DJHC flames using tabulated chemistry and stochastic combustion model. The prime focus of their study was to understand the role of entrainment on flame stabilization and the effects of Reynolds number on the flame lift-off height. They reported that the LES could predict the lift-off height reasonably well but the temperature profiles displayed inconsistency between the predictions and the experimental measurements at the jet exit. More recently, De et al. [16] studied these flames using both MEPDF and Lagrangian PDF (LPDF) methodologies. They found that both MEPDF and LPDF predictions were in good agreement excluding some discrepancies at the downstream region. The limitations of the turbulence and combustion models to accurately model the JHC flames are primarily due to the inaccuracies in modeling turbulence-chemistry interaction, which is one of the major sources of errors while modeling turbulent reacting flows.

From the above discussion, it is obvious that the non-linear interaction between fluid mixing and finite rate chemistry in the MILD regime makes it a challenging task for the combustion community. In the past three decades, a number of approaches have been devised to calculate the mean reaction source term using turbulence-chemistry interaction. The transported PDF methods can be considered as one of the most elaborate and accurate method to model the turbulent combustion as the reaction source terms are in the closed form and require no modeling. A detailed description of the transported PDF methods can be found in Pope [24] and Fox [25] whereas a complete review on the transported PDF methods is available in Haworth [26]. The Direct-Quadrature-Methods-of-Moments (DQMOM) [27] scheme was first put forth by Fox [25]. This method involves the solution of transport equations for weights and abscissa of quadrature points. This method can be used alternatively to solve the transported PDF equations after certain approximations [25]. Initially, Fox et al. [28, 29] used this approach to model the population balance equation in order to evaluate the PDF for particle size but later Fox [25] extended this scheme to model the turbulent combustion by considering a multi-environment presumed shape PDF. This presumed shape PDF consists of a series of delta functions which are, then, incorporated into the joint composition PDF transport equation which, in-turn, results into a set of transport equations relating the weights and abscissa of the quadrature points. Later, the Interaction-by-Exchange-with-the-Mean (IEM) [30] mixing model was incorporated in this formulation to provide closure for the micro-mixing term. This scheme was, altogether, known as DQMOM-IEM or Multi-Environment Eulerian PDF (MEPDF) model. In the MEPDF model, the reaction source term is obtained in the closed form in the transport equations which are then solved in the Eulerian frame. Therefore, the MEPDF model retains the advantage of PDF transport model but is free from the statistical errors. Also, as per studies carried out by Tang et al. [31], Akroyd et al.[32-33], and Rakesh et al. [34] emphasized that a small number of environments is sufficient for reasonable accuracy which makes this method computationally efficient. Several other studies [35- 36] have been performed with MEPDF model to understand its applicability and efficiency in modeling turbulence-chemistry interaction but none of them reported the use of this methodology to study MILD combustion.

Hence, the primary focus of the present work is to study MILD combustion using MEPDF methodology in order to precisely understand the behavior of this particular model. 2D RANS predictions obtained using the MEPDF model, for both the burners, are compared with the respective experimental database for the corresponding Reynolds number i.e. Re=4100 and Re=8800 for the DJHC burner and Re=10000 for the Adelaide burner.

## 2. MULTI ENVIRONMENT EULERIAN PDF TRANSPORT

The PDF transport modeling starts with the joint-composition PDF transport equation given as follows;

$$\frac{\partial \rho f_\varphi}{\partial t} + \frac{\partial}{\partial x_i}\left[\rho u_i f_\varphi\right] + \frac{\partial}{\partial \psi_k}\left[\rho S_k f_\varphi\right] = -\frac{\partial}{\partial x_i}\left[\rho \left\langle u_i'' | \psi \right\rangle f_\varphi\right] + \frac{\partial}{\partial \psi_k}\left[\rho \left\langle \frac{1}{\rho}\frac{\partial J_{i,k}}{\partial x_i} | \psi \right\rangle f_\varphi\right] \quad (1)$$

where $f_\phi$ is the single-point, joint probability density function (PDF) of species composition and enthalpy. The unsteady rate of change of PDF is represented by the first term on the left hand side in eqn.(1) whereas the second and third term represents the transport in physical space by mean velocity and transport in composition space by chemical reactions, respectively. While, on the right hand side the first term represents the transport due to velocity fluctuations whereas the molecular mixing is represented by the second term. The right hand side of the eqn. (1) contains unclosed terms where the first term is closed by a turbulence model and the second term is closed by a micro-mixing model.

As explained earlier, in the MEPDF method, the closed form joint composition PDF transport equation is approximated using a presumed shape PDF which consists of a series of delta functions [25]. Using this approach, the eqn. (1) can be represented by a collection of $N_e$ delta functions given below;

$$f_\phi(\psi;x,t) = \sum_{n=1}^{N_e} w_n(x,t) \prod_{\alpha=1}^{N_s} \partial[\psi_\alpha - <\phi_\alpha>_n (x,t)] \tag{2}$$

where $N_s$ is the number of species, $N_e$ is the number of environments, $w_n$ is the weight (or probability) of each environment, and $<\phi_\alpha>_n$ is the mean composition vector in the $n^{th}$ environment. By substituting this definition of presumed shape PDF (eqn. 2) into the composition PDF transport equation (eqn. 1), we obtain the following set of transport equations for any environment.

$$\frac{\partial \rho p_n}{\partial t} + \frac{\partial}{\partial x_i}(\rho u_i p_n) = \Gamma \nabla^2 p_n \tag{3}$$

$$\frac{\partial \rho \vec{s_n}}{\partial t} + \frac{\partial}{\partial x_i}(\rho u_i \vec{s_n}) = \Gamma \nabla^2 \vec{s_n} + C_\phi \frac{\varepsilon}{k}(\langle\phi_i\rangle - \langle\phi_i\rangle_n) + p_n S(\langle\vec{\phi_n}\rangle) + \vec{b_n} \tag{4}$$

where, $\vec{s_n} = p_n \vec{Y_n}$ or $p_n H_n$

The eqn.(3) represents the transport of the probability of occurrence of $n^{th}$ environment while the eqn.(4) represents the transport equation for the probability of weighted species mass fraction or probability weighted enthalpy in each environment. In eqn.(4), on the right hand side, the second term represents the micro-mixing term which is closed using the IEM closure. The third term represents the reaction source term whereas the last term ($b_n$) is known as the correction term which takes care of the modeling assumptions used in the MEPDF model. These correction terms are approximated as;

$$\sum_{n=1}^{N_e} \langle\phi\rangle_n^{m_j-1} b_n = (m_j - 1) \sum_{n=1}^{N_e} \langle\phi\rangle_n^{m_j-2} p_n c_n \tag{5}$$

In the MEPDF formulation, eqn.(4) is obtained by taking moments of the PDF transport equation and the correction terms are obtained by forcing the eqn.(4) to match the moments of the PDF transport equation. Fox and Wang [37] have highlighted the issue of boundedness and singularity of the co-variance matrix of different scalars in calculating generic solutions to find all the correction terms. These issues are handled by imposing some added constraints, like, ignoring cross moments and handling each scalar independently. These two constraints result in a simplified form of correction terms given by eqn.(5). The DQMOM method is applied to derive (eqn.5) the correction terms. The derivation begins by inserting eqn.(2) into eqn.(1) and following a derivation explained in Fox [25], we obtain a set of linear equations, eqn.(3) and eqn.(4), for $N_s$ scalars. The $N_e(N_s+1)$ unknowns are calculated by solving eqn.(4) with lower order moments. The zero$^{th}$ and first order moments must be satisfied and yield $(N_s+1)$ linear equations whereas the remaining $(N_e-1)(N_s+1)$ linear equations must be calculated from higher order moments. The entity '$m_j$' in equation (5) represents the order of the moment whose value varies as $m_j=1,2,....,N_e$. Here $N_e$ represents the number of environments whereas $N_s$ represent the number of scalars considered. The value of $m_j$ depends on the number of environments considered. The detailed derivations of eqn. (2) to eqn. (5) along with underlying assumptions can be found in Fox [25].

These governing transport equations are discretized using the finite volume method and are solved for each environment. A stiff ODE solver is used to calculate the reaction source term for the each environment whereas the In-Situ-Adaptive-Tabulation (ISAT) [38] is used to tabulate the source terms. In the present study, we have used a two environment formulation to solve the joint-composition PDF transport equation. The probability of environment-1 is taken as 1 for fuel jet and 0 for coflow whereas that of environment-2 has been set to 1 for coflow and 0 for fuel jet. Therefore, environment-1 has weight leaning to fuel jet while environmenr-2 has weight leaning towards coflow stream. Thus, environment-1 affects the physical processes in the fuel rich side on composition space while environment-2 dominates the processes in the fuel lean side. The micro-mixing model couples and represents the evolution of different scalars in both the environments.

Previous studies [5-16] concerning MILD combustion are all 2D studies. RANS based turbulence models give averaged values and not instantaneous values therefore undertaking a 3D study won't produce improved results unless a more accurate turbulence models like Large Eddy Simulation (LES) based turbulence models are used. Since one of the objectives of this study is to assess the accuracy of RANS based turbulence model in conjunction with PDF based combustion models, a 2D simulation approach has been used in the present paper.

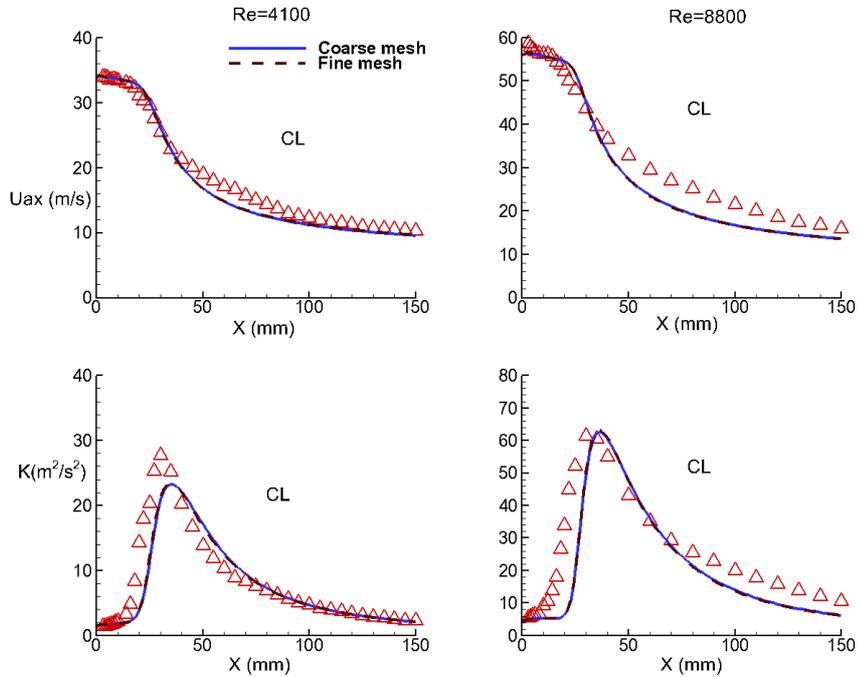

Figure 1: Centerline profiles for mean axial velocity and turbulent kinetic energy obtained for DJHC-I flames obtained using coarse and fine mesh. Symbols are experimental measurements and lines are predictions.

## 3. DELFT-JET-IN-HOT-COFLOW (DJHC) BURNER

In this section, the MEPDF predictions for DJHC burner are reported. The current computations are done with only two environments. A parametric study of the MEPDF model is also performed.

3.1 Burner description and numerical set-up

DJHC burner was designed in the TU-Delft and experimental database is available for modeling [2]. The burner consists of a central fuel jet pipe with internal diameter of 4.5mm which is surrounded by an outer tube of internal diameter 82.8mm. The hot coflow is generated by the secondary burner situated inside the outer tube. More details on the DJHC burner can be found in literature [2-4, 14].

Since the burner is symmetric, a 2D axisymmetric grid is used in the current work. In the experiments carried out by Oldenhof et al. [2-4], the first LDA measurement locations were mentioned at z=3mm from fuel exit. Considering this, the computational domain starts 3mm downstream of the jet exit. The domain extends for 225mm in the axial direction whereas in the radial direction it extends for 80mm which accounts for the entrainment of ambient cold air. The current computations are carried out using ANSYS Fluent 13.0 [39]. The convective fluxes of all the equations were discretized using second order upwind scheme. Pressure and velocity are coupled using the SIMPLE algorithm. In order to model turbulence, the RANS based realizable k-ε (RKE) turbulence model is used. The micro-mixing term in the eqn. (4) is modeled using the IEM mixing model. The chemistry is modeled using a reduced mechanism, DRM 19 [21], involving 19 species and 84 reactions. Simulations have been performed with the Dutch Natural gas in the central fuel jet for varying $O_2$ percentage, i.e. 7.6 and 10.9 %, in the coflow. The mean temperature, mean velocity applied at the inlet boundaries of both fuel jet and coflow are taken from the experimental database [2]. The experimental values of radial profiles at x=3 mm are used to define the boundary conditions of the physical quantities at both fuel and coflow inlets.

The species concentrations are calculated using the equilibrium assumption whereas the turbulent kinetic energy and dissipation are calculated from the measured axial and radial normal components of the Reynolds stresses by assuming the azimuthal component to be equal to radial one. More details on the numerical description can be found in the previously published work by De et al. [14].

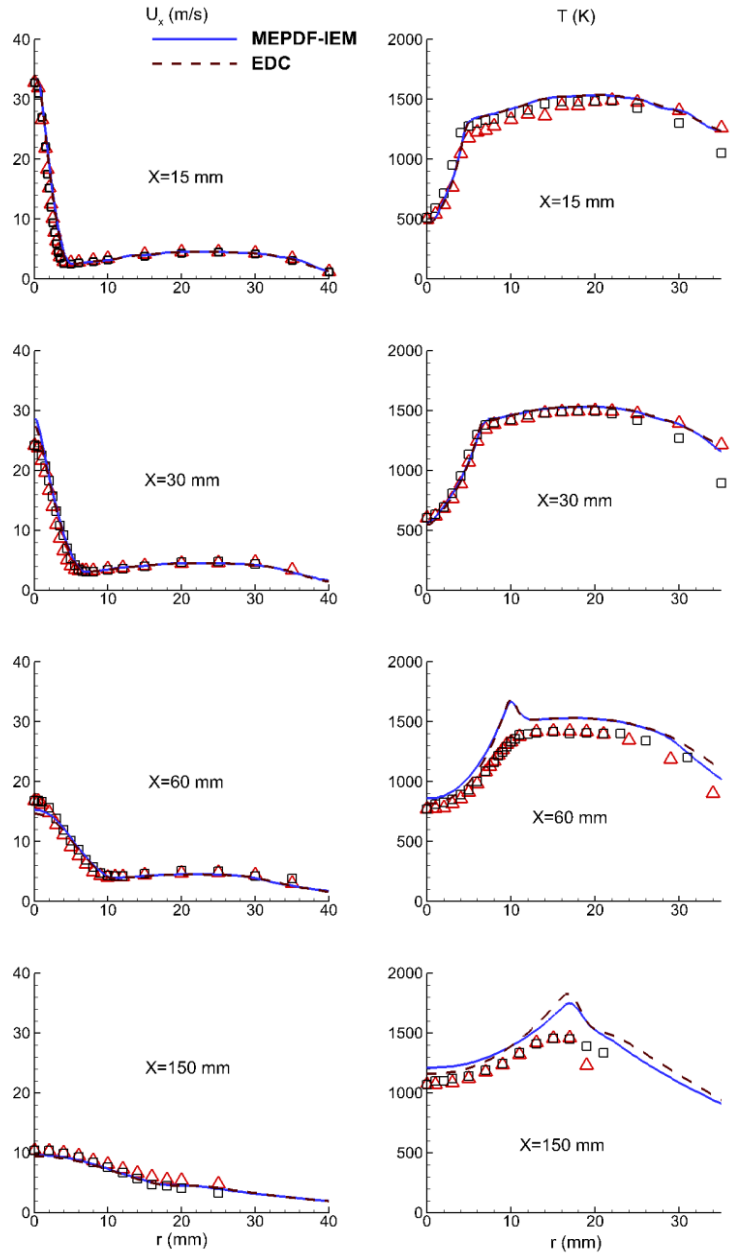

Figure 2: Radial profiles of mean axial velocity and mean temperature obtained for DJHC-I flames (Re=4100) with $C_\varphi$ =2. Symbols (*triangle* 0 ≤ r ≤ 35, *squares* -35 ≤ r ≤ 0) are experimental measurements and lines are predictions.

3.2 Results and discussion

The results obtained after performing detailed simulations for the DJHC flames, exhibiting the coflow properties explained in Oldenhof et al.[2], at Reynolds number Re=4100 and Re=8800, are discussed in this section. First, the predictions for the mean axial velocity, turbulent kinetic energy, mean temperature, and Reynolds stress components obtained using MEPDF model for Re=4100 and Re=8800 are discussed. The current results are compared with the experimental database provided by Oldenhof et al.[2] where the oxygen concentration has been kept constant at 7.6% (DJHC-I flames) in both cases reported here. The study is further extended for the predictions of the DJHC-X flames with excess oxygen (10.9% $O_2$) with Re=4600 [2, 14].

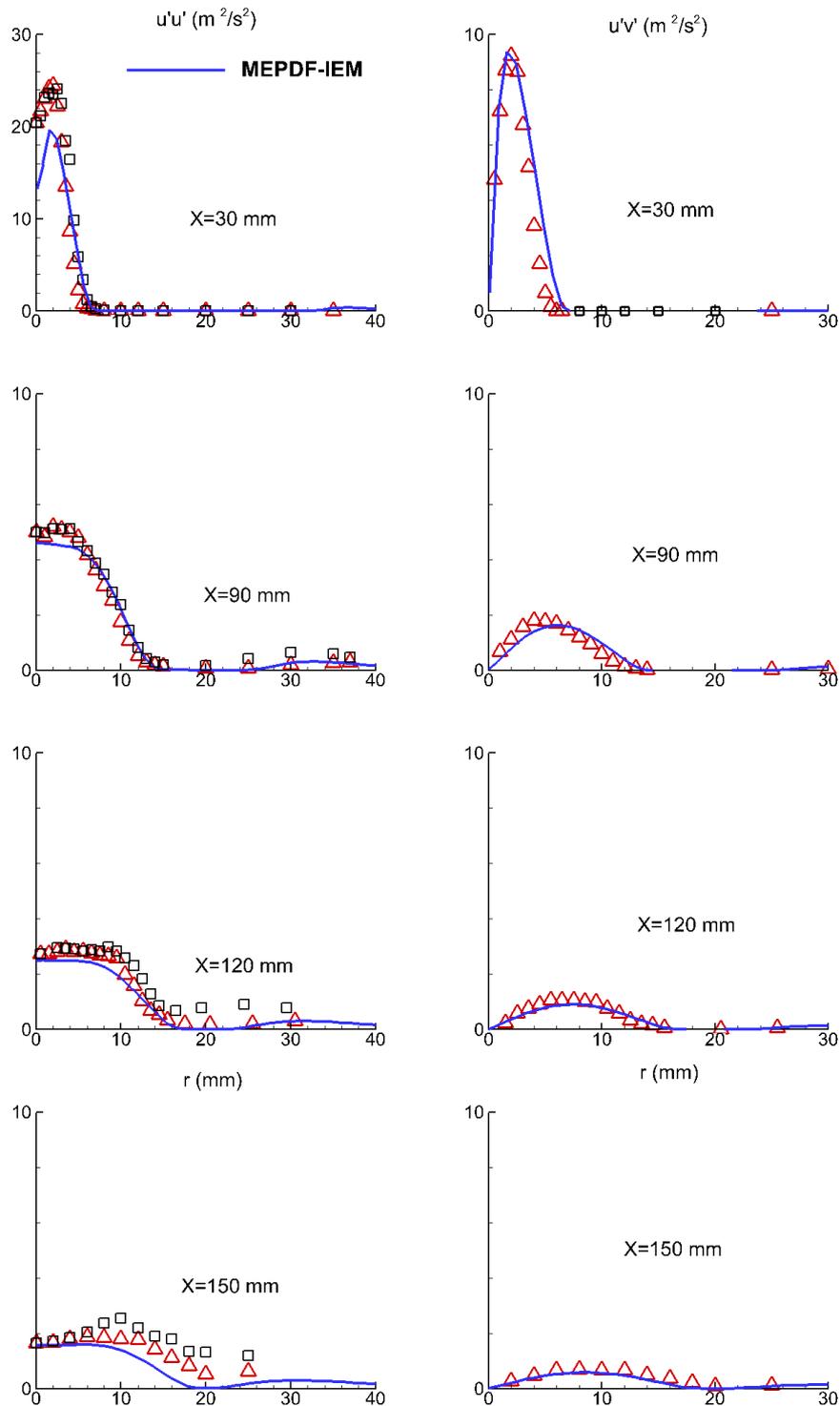

Figure 3: Radial profiles of Reynolds stresses obtained for DJHC-I flames (Re=4100) with $C_\varphi$=2. Symbols (*triangle* 0 ≤ r ≤ 35, *squares* -35 ≤ r ≤ 0) are experimental measurements and lines are predictions.

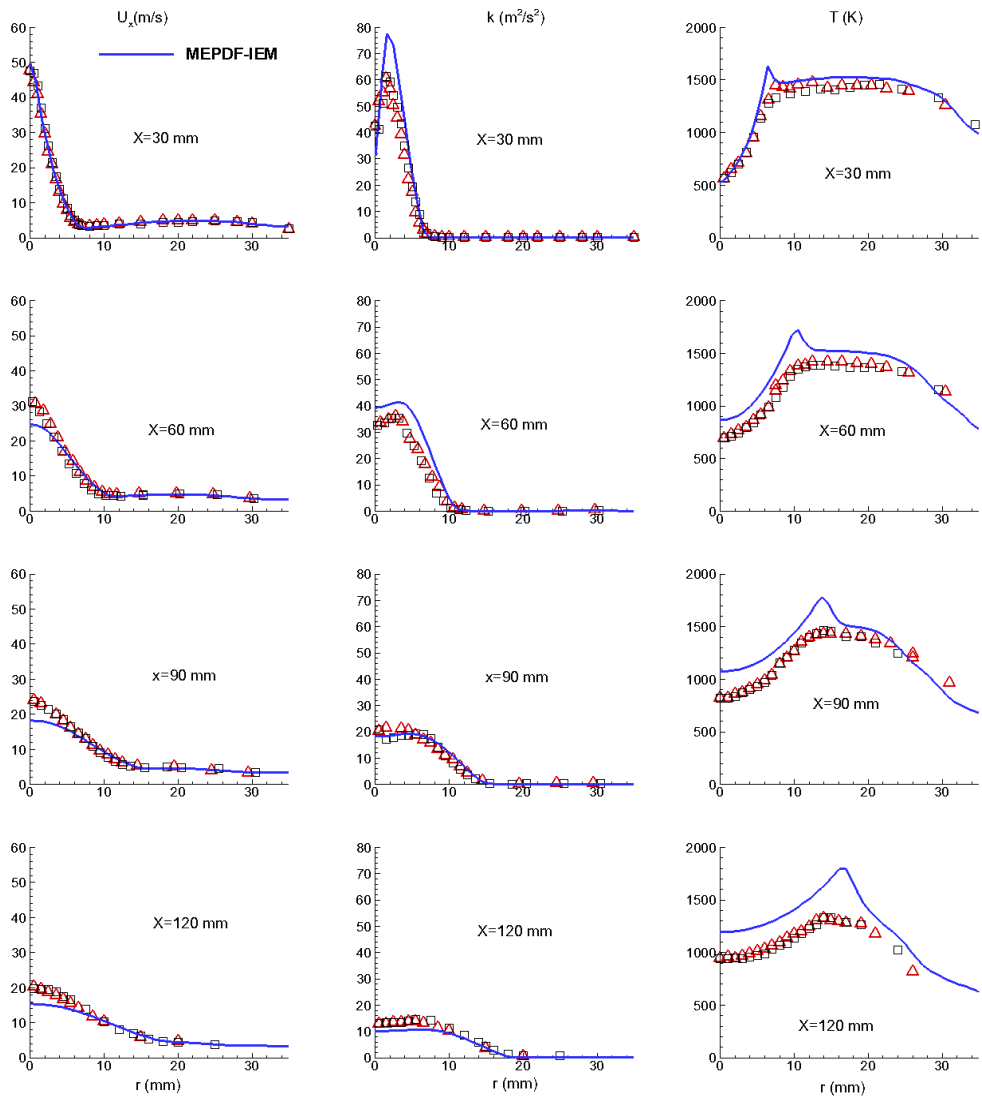

Figure 4: Radial profiles of mean axial velocity, turbulent kinetic energy, and mean temperature obtained for DJHC-I flames (Re=8800) with $C_\varphi$=2. Symbols (*triangle* 0 ≤ r ≤ 35, *squares* -35 ≤ r ≤ 0) are experimental measurements and lines are predictions.

A grid independence study is carried out using two different grids for the cases with lower (Re=4100) as well as higher Reynolds number (Re=8800). The first grid used is a coarse one with 180x125 (axial x radial) grid points whereas the finer grid consists of 360x250 grid points. The predictions using both the meshes are found to be in good agreement with each other as reported in the literature [16]. Hence, we report the detailed results using the coarse mesh only in the next sub-sections.

Figure 1 depicts the center line mean axial velocity and turbulent kinetic energy plots obtained using the MEPDF combustion model along with RKE turbulence model and DRM 19 [21] chemical mechanism. As observed from the figure, the predictions exhibit proper trends of mean profiles, and are in good agreement with the experimental results. Figure 2 shows the radial profiles of mean axial velocity and temperature for the DJHC-I flames (Re=4100) obtained using MEPDF-IEM. The results obtained using the EDC model by De et al. [14] are plotted alongside the MEPDF predictions for better comparison. It is observed that the mean axial velocity profile is nicely captured by the MEPDF model while the center line values are slightly under-predicted in the downstream region. Whereas in the case of mean temperature profiles, it is properly predicted till axial location x=30 mm and after which a temperature peak at x=60 mm is observed that indicates a sign of early ignition. In the downstream region, the temperature profiles are over-predicted. The peak temperature obtained through the MEPDF model is approximately 1700K which is 13% higher than the experimental observation [2]. The peak temperature is observed at radial location of r=17mm at x=150mm. Comparing the current peak temperature profiles to those obtained in De et al. [14], MEPDF predictions are found to be on the lower side as the EDC predictions exhibit a peak temperature of approximately 1830K which is 25% higher than the measurements at the same radial location mentioned above.

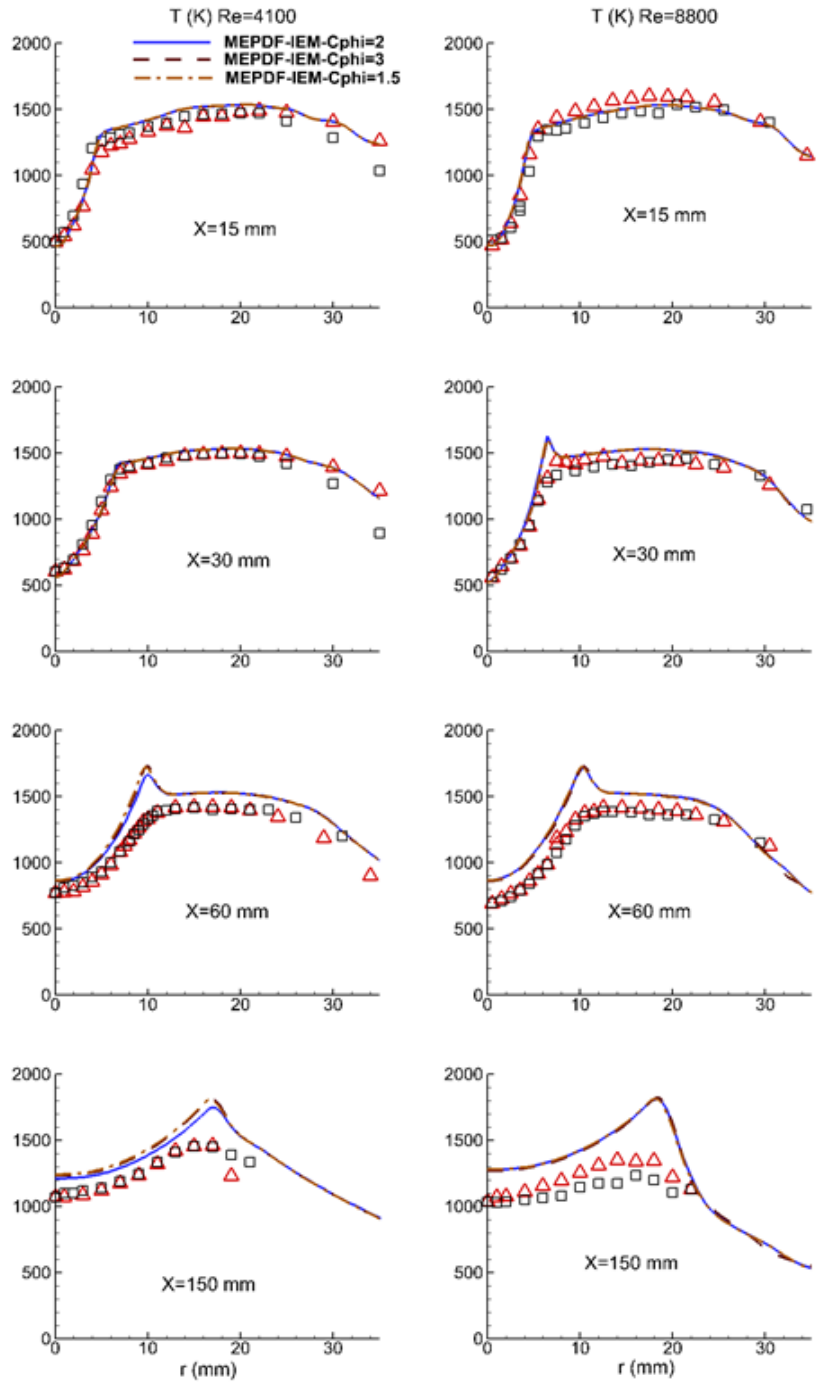

Figure 5: Radial profiles of mean temperature obtained for DJHC-I flames (Re=4100 and Re=8800) for $C_\varphi$ comparison. Symbols (*triangle* 0 ≤ r ≤ 35, *squares* -35 ≤ r ≤ 0) are experimental measurements and lines are predictions.

This temperature rise in the shear layer is due to mixing between the hot coflow and the fuel jet. Thus, the accuracy of the temperature predictions will depend on the accuracy of the mixing model, both macro-mixing and the micro (or molecular) mixing. In the present formulation, we have used the IEM approach for modeling the micro-mixing term in eqn. (4). In case of IEM, the composition of all the scalars in an environment relaxes towards the mean composition at the same rate. Therefore, the effect of localness in the flow field is missing which can be a potential source of error. The correction terms of the PDF transport equation (eqn.1) can be another source of errors. Moreover, temperature is highly non-linear function of mixture fraction in the high temperature zone around the stoichiometric mixture fraction. Therefore, large discrepancies are observed as we move away from the jet exit towards the flame stabilization in the downstream region. Therefore, the inaccuracies in the IEM micro-mixing model, presence of the fuel and oxidizer in stoichiometric proportions can possibly cause the temperature field to shoot up.

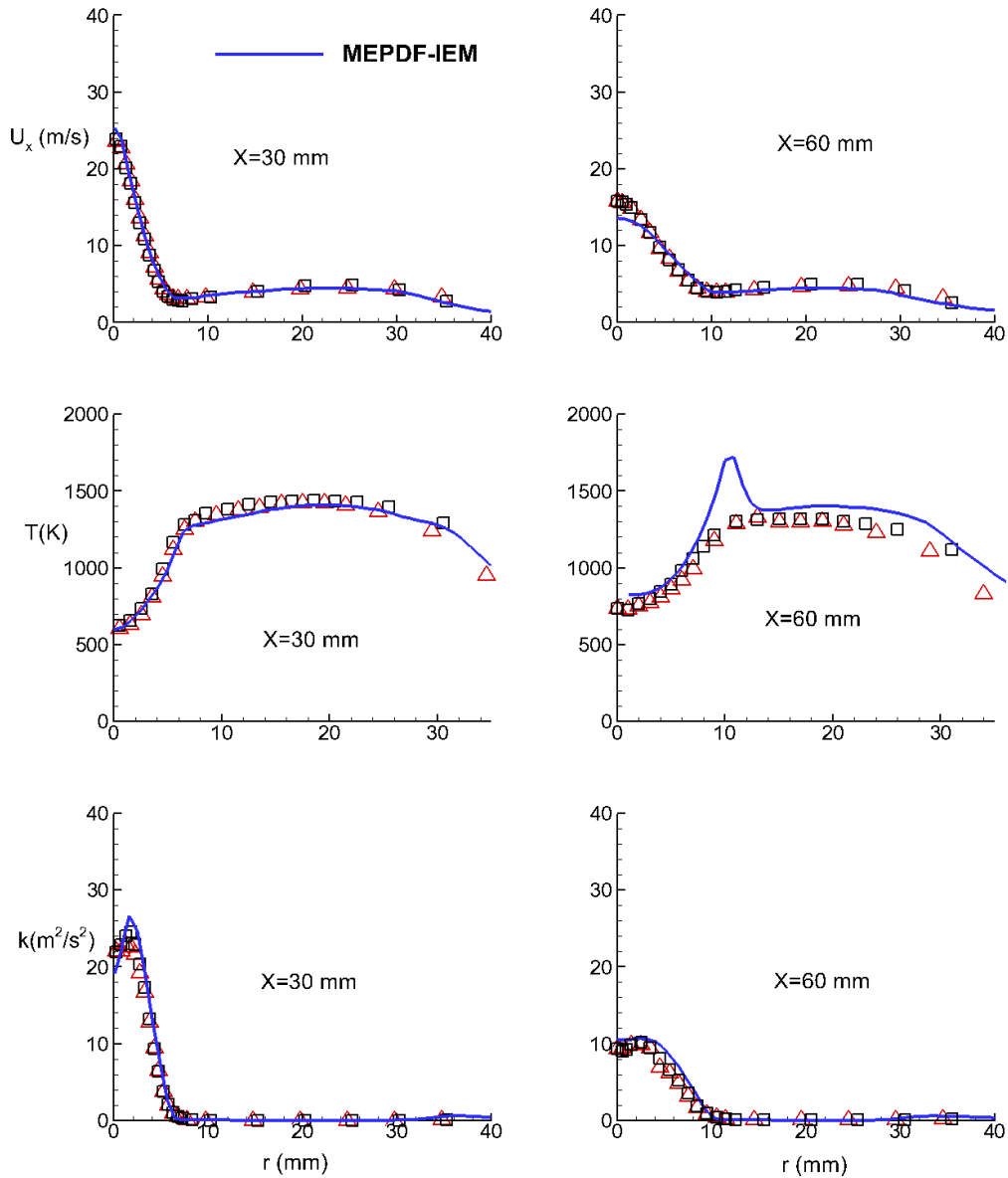

Figure 6: Radial profiles of mean axial velocity, temperature, and turbulent kinetic energy obtained for DJHC-X (Re=4600) with $C_\varphi$=2. Symbols (*triangle* 0 ≤ r ≤ 35, *squares* -35 ≤ r ≤ 0) are experimental measurements and lines are predictions.

Figure 3 shows the radial profiles of the Reynolds stresses obtained for the DJHC-I flames. The Reynolds stress is an important quantity which describes the mixing of momentum between fuel jet and coflow and is obtained using the Boussinesq hypothesis. As observed from the Fig. 3, the normal stress component (u'u') is under-predicted at axial location x=30 mm after which the profiles are nicely captured by the model in the downstream region whereas the Reynolds shear stress (u'v') is found to be in good agreement with the experimental results. The discrepancies observed in predicting Reynolds stresses can be attributed to the accuracy of the turbulence model. The combined effect is clearly visible in the center line profiles of turbulent kinetic energy (Fig. 1).

The influence of turbulence-chemistry interaction on the flame characteristics can be increased by increasing the Reynolds number which serves as a better case to study the applicability of the MEPDF model to simulate flames featuring strong turbulence-chemistry interaction. Therefore, the calculations are repeated here for the DJHC-I flames by increasing the Reynolds number to 8800.

The radial profiles of the mean axial velocity, turbulent kinetic energy, and mean temperature obtained for the DJHC-I flames (Re=8800) are reported in Figure 4. The mean axial velocity profiles are accurately captured at the initial locations after which the profiles are nicely captured with a slight under-prediction. We obtained a similar trend for the lower Reynolds number case (Re=4100) but the under-prediction further magnifies in the higher Reynolds number case (Re=8800). This under-prediction of the axial velocity profiles can be attributed to the over-estimation of the turbulent viscosity which increases the diffusion rate of the jet due to which the shear layer spreads radially and, thus, the jet spreads more in the radial direction which results in the under-estimation of the velocity profiles. From the figure, we can, also observe that the turbulent kinetic energy is over-predicted till axial location x=60 mm after which it is nicely captured by the model. The discrepancies observed can be attributed to the inaccuracies in the turbulence model.

When we compare the trends obtained for the mean temperature in low (Re=4100) (Fig.2) as well as high (Re=8800) (Fig.4) Reynolds number case, we can observe that a similar trend is observed where temperature profiles are over-predicted in the downstream region. The over-prediction increases with increase in Reynolds number. A similar methodology, which was explained in the lower Reynolds number case (Re=4100), can be used to explain the behavior of the results obtained. But comparing the Figs. 2 and 4, we can observe that the sharp peak in temperature first observed at x=60 mm for Re=4100 case, while it has now been shifted to x=30 mm for the higher Reynolds number (Re=8800) case due to faster evolution of scalars in different environments. One way to possibly improve the predictions is to use a more detailed chemical mechanism or better mixing model in order to handle the kinetically controlled reactions in MILD regime.

While looking at the temperature and velocity field, the predictions are little misleading. The over-prediction in temperature field may have some errors in density field but on the other hand, the velocity profiles are nicely predicted. The reason is that the density field has a collective effect of temperature and species mass fraction and affects only the convective terms whereas the velocity profiles have a combined effect of convective as well as diffusive terms which are governed by the turbulence model. Therefore, the trend obtained in velocity and temperature profiles need not to be identical.

3.2.1 Effects of $C_\varphi$

Comparing the overall results obtained using the lower as well as the higher Reynolds number, we can say that the accuracy of the MEPDF model decreases as we increase the Reynolds number from Re=4100 to Re=8800. This can be attributed to the inaccuracies in the turbulence model and an inappropriate response of the MEPDF model through micro-mixing term. The present MEPDF formulation can be improved by increasing the number of environments but it is always not a viable option since increasing the number of environments will increase the cost of computation without yielding significant improvements to the predictions as observed in the present scenario. Furthermore, this becomes against the attractive features of efficient Eulerian PDF methodology to give reasonably good results with less number of environments. Another way to improve the results is to adjust the micro-mixing term. In context of MEPDF model, the micro-mixing term can be controlled by changing the value of mixing constant, i.e. $C_\varphi$. $C_\varphi$ can directly affect the preparation of the combustible mixture, and hence, can delay or advance the initiation of the reaction. In this case, since we are studying the MILD combustion which features slow reaction rates, it is desirable to delay the reaction. Therefore, a lower value of $C_\varphi$ is suitable. But care must be taken while selecting the value of $C_\varphi$, since reducing it beyond a certain limit can lead to global extinction. In order to study the effect of change in mixing constant $C_\varphi$, a comparative study is carried out for both the Reynolds number (Re=4100 and Re=8800) cases by varying $C_\varphi$ value between 1.5, 2, and 3. Figure 5 shows the radial profiles of the mean temperature for Re=4100 and Re=8800 case respectively. From the figure, we can observe that changing the value of $C_\varphi$ to 1.5 can improve the results slightly but only in the near-jet exit area till x=60 mm after which the profiles are over-predicted compared to what we obtained using $C_\varphi$=2 and make no substantial changes in the results. For the Re=4100 case, varying the $C_\varphi$ value from 1.5 to 3 leads to approximately 50K rise in temperature whereas no substantial change in temperature is obtained for the Re=8800 case. Therefore, we can conclude that this flame is not very sensitive to change in $C_\varphi$ and hence the sources of error are more due to the chemistry.

3.2.2 Effects of $O_2$ concentration (Re=4600, 10.9% $O_2$)

In order to understand the effect of different oxygen concentration in the coflow on the flame characteristics and the sensitivity of the MEPDF approach to predict the change in flame structure with change in O2 concentration, we now report the results obtained for the DJHC-X flames with excess oxygen (10.9% $O_2$) in the coflow. Figure 6 shows the radial profiles of the mean axial velocity, mean temperature, and turbulent kinetic energy obtained for DJHC-X flames at axial locations x=30mm and x=60mm. The model is able to accurately capture the mean axial velocity profiles initially after which we can observe slight under-prediction in the inner shear layer in the downstream region (x=60mm). Comparing Figs. 6 and 2, we can observe that a similar trend in the temperature profiles is obtained in the near-jet exit area but as we move downstream, the excess oxygen case (10.9% $O_2$) significantly over-predicts the flame temperature as compared to the 7.6% $O_2$ case. The turbulent kinetic energy profiles, as observed from the figure, are slightly over-predicted in the inner shear layer but are accurately captured in the outer shear layer. These discrepancies observed in predicting the turbulent kinetic energy along the inner shear layer can be associated with the inaccuracies of turbulence model in which the production and dissipation of kinetic energy is not properly modeled. Overall, the MEPDF model produces agreeable profiles for the excess oxygen case while comparing the results with experimental database except the peak temperature profiles in a narrow radial zone, which are over-predicted in the downstream region.

3.2.3 Flame lift-off height

As per the literature by Oldenhof et al. [2], flame pockets at their first instant of appearance are referred to as ignition kernels. These ignition kernels then convect downstream to form a stable flame. The flame lift-off height is defined as the height at which the first ignition kernel is detected. OH mass fraction is used as the means to detect the ignition kernel. An ignition kernel is registered when the OH mass fraction reaches a value of 1e-03. A similar methodology has been used here to determine the flame lift-off height. The predicted lift-off height is observed at ~ 33mm for the Re=4100 case while ~24 mm for Re=8800 case. Similar data was also observed in the EDC predictions as reported by De et al. [14]. The MEPDF model quite significantly under-predicts the lift-off height for both the Reynolds number cases as the experimentally measured lift-off height was reported at ~84mm.

　The lift-off height, for the JHC flames, depends on the entrainment of hot coflow into the fuel stream. At a higher Reynolds number, since the fuel stream velocity is higher compared to the low Reynolds number case, better entrainment is obtained for this case which results in establishment of the flame at a lower axial location as compared to the low Re number case, thus, a lower lift-off height. In low Re number case, since the fuel jet velocity is lower, entrainment of hot coflow is reduced compared to the higher Re case which increases the lift-off height as the flame is now established at a higher axial location. Even though, the model is able to capture the effect, it under-predicts the lift-off height compared to the experimental value. The micro-mixing model couples and represents the evolution of different scalars in both the environments, in this case, the fuel stream and the hot coflow. Thus, the micro-mixing model is responsible for mixing between the environments as it brings together fuel stream and the hot coflow. The MEPDF approach accurately predicts the trend but the predicted lift off heights are shorter. This is primarily due to the inadequate response of the IEM model.

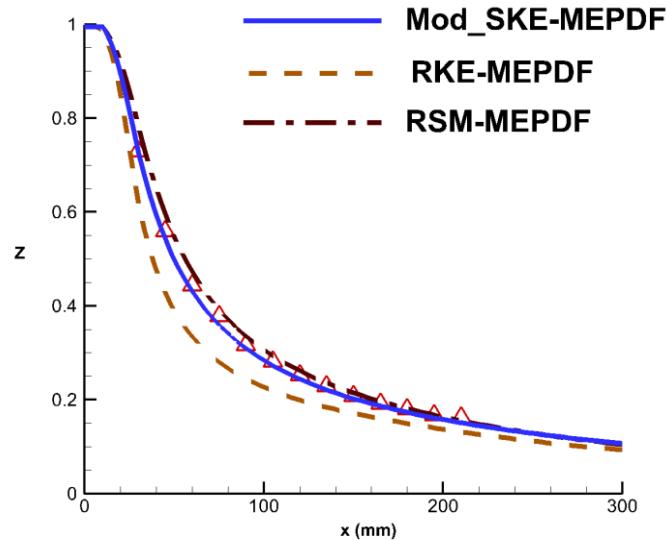

Figure 7: Centerline profiles of mean mixture fraction obtained for HM1 flame obtained using different turbulence models. Symbols are experimental measurements and lines are predictions.

| | Fuel Jet $(CH_4$-$H_2)$ | | Oxidant (Coflow) | | | | |
|---|---|---|---|---|---|---|---|
| Flame | Re | T (K) | T (K) | Y ($O_2$)% | Y ($N_2$)% | Y($H_2O$)% | Y($CO_2$)% |
| HM1 | 10000 | 305 | 1300 | 3 | 85 | 6.5 | 5.5 |

Table 1: Detailed boundary conditions for HM1 flames.

## 4. ADELAIDE JHC BURNER

In case of DJHC burner, no species data is available. In order to have detailed understanding of the predictive capabilities of MEPDF model, it is necessary to consider some other burner. That's why we have considered the Adelaide burner and discuss it in the current section which provides an extensive database of species profiles.

4.1 Burner description

The main difference between DJHC burner and Adelaide burner is that $N_2$ is used to cool down the central fuel jet in the Adelaide burner to that of convective and radiative heat losses along the central fuel jet in the DJHC burner. Another striking difference between the two burners is the design of the secondary burner. In the DJHC burner, the design of the secondary burner allows the addition of the seeding particles, to be used as tracers in the LDA or PIV measurements, whereas the Adelaide burner has no such arrangement for particle seeding. The Adelaide burner consists of an insulated and cooled central fuel jet with an internal diameter of 4.25mm and an outer annulus of an internal diameter 82mm. A cold mixture of air and nitrogen, which helps to create a hot coflow through the secondary burner, also helps to cool down the secondary burner which is mounted upstream of the jet exit. To minimize the heat loss to the surroundings, the outer annulus is insulated using ceramic straps. More details and schematics of the Adelaide burner can be referred from Dally et al. [1].

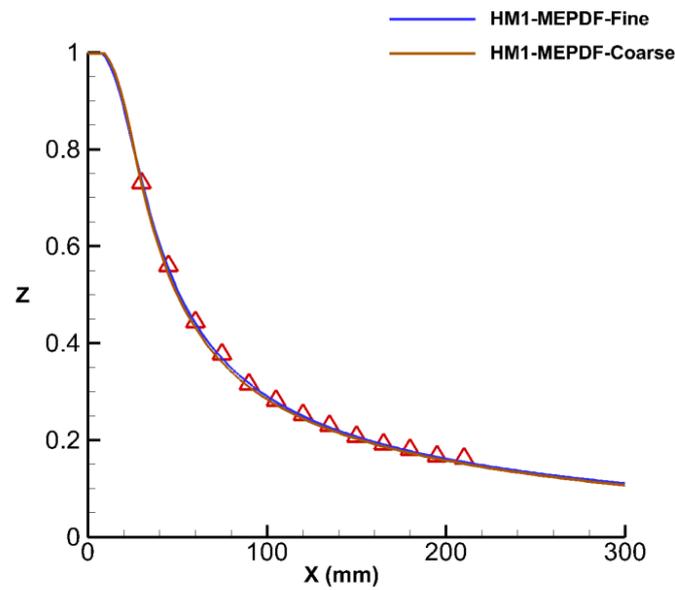

Figure 8: Centerline profiles of mean mixture fraction using two different grids for HM1 flames at Re=10000. Symbols are experimental measurements and lines are predictions.

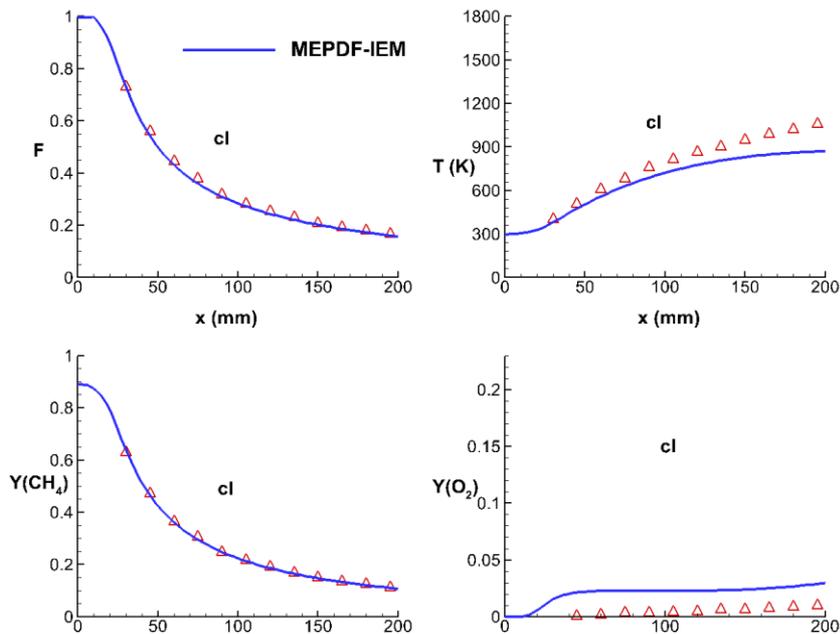

Figure 9: Centerline profiles of mean mixture fraction, temperature, $CH_4$ and $O_2$ mass fraction for HM1 flames with $C_\varphi=2$. Symbols are experimental measurements and lines are predictions.

4.2 Numerical set-up and boundary condition

A similar numerical set-up, as explained in the previous section, with a few changes has been used to simulate the methane-hydrogen ($CH_4$-$H_2$) JHC flames in the Adelaide burner. The HM1 flame, as per the experiments, with oxygen mass fraction of 3% (by mass) in the hot coflow has been simulated using ANSYS Fluent 13.0 [39]. Considering the symmetry of the burner, a 2D axi-symmetric grid, similar to DJHC burner but with a few changes in the dimensions, is used in this case. In the Adelaide burner, the central fuel jet protrudes at the jet exit by 4mm, and hence, the computational domain starts 4mm downstream of the jet exit and extends for 300mm in the axial direction and 80mm in the radial direction. A modified SKE turbulence model ($C_{\varepsilon1}=1.6$) is used to model the turbulence whereas IEM mixing model is used to close the micro-mixing term in the eqn.(4). DRM 19 [21] chemical mechanism has been used to describe the chemistry.

In the axial direction, at the end of the computational domain, the boundary condition is set to outflow. The inlet boundary conditions (fuel jet and coflow) for temperature and species mass fractions are referred from the experimental database [1] as provided in Table 1. The simulated fuel jet Reynolds number is Re=10000. The velocity inlet boundary conditions for the hot and cold coflow are set to 3.2 m/s and 3.3 m/s respectively. It is observed from the literature that the solution is highly sensitive to the turbulent quantities at the inlet, hence, the turbulent intensity at the hot and cold coflow are set to 5% whereas for the fuel jet it was set to 7% based on the published literature [8]. For the Adelaide burner, we tested RKE and RSM turbulence models. In Frassoldati et al. [8], the authors had tested effects of various inflow conditions including turbulence models. They reported that SKE model with a modified dissipative coefficient ($C_{\varepsilon1}=1.6$) produced the best results for the Adelaide JHC flames. We have carried out similar simulation with modified SKE turbulence model and compared with results obtained using RKE and RSM model and found better results with modified SKE model compared to other turbulence models as shown in figure 7. Therefore, we have used the modified SKE model to model turbulence in this case. The detailed boundary conditions for fuel stream and hot coflow are given in table 1.

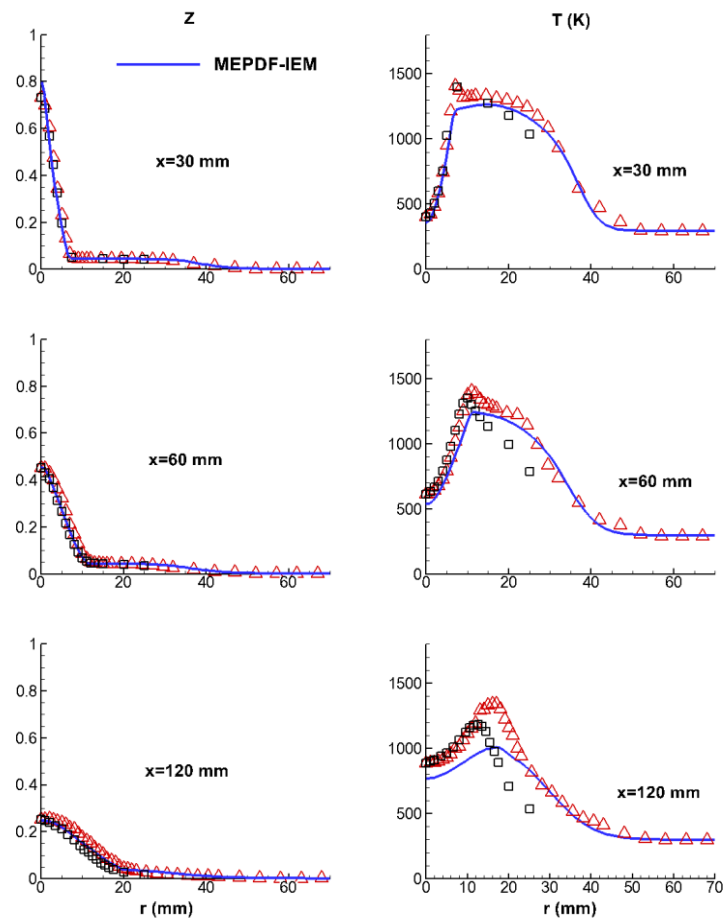

Figure 10: Radial profiles of mean mixture fraction and mean temperature for HM1 flames with $C_\varphi$=2. Symbols (*triangle* 0 ≤ r ≤ 70, *square* -25 ≤ r ≤ 0) are measurements and lines are predictions.

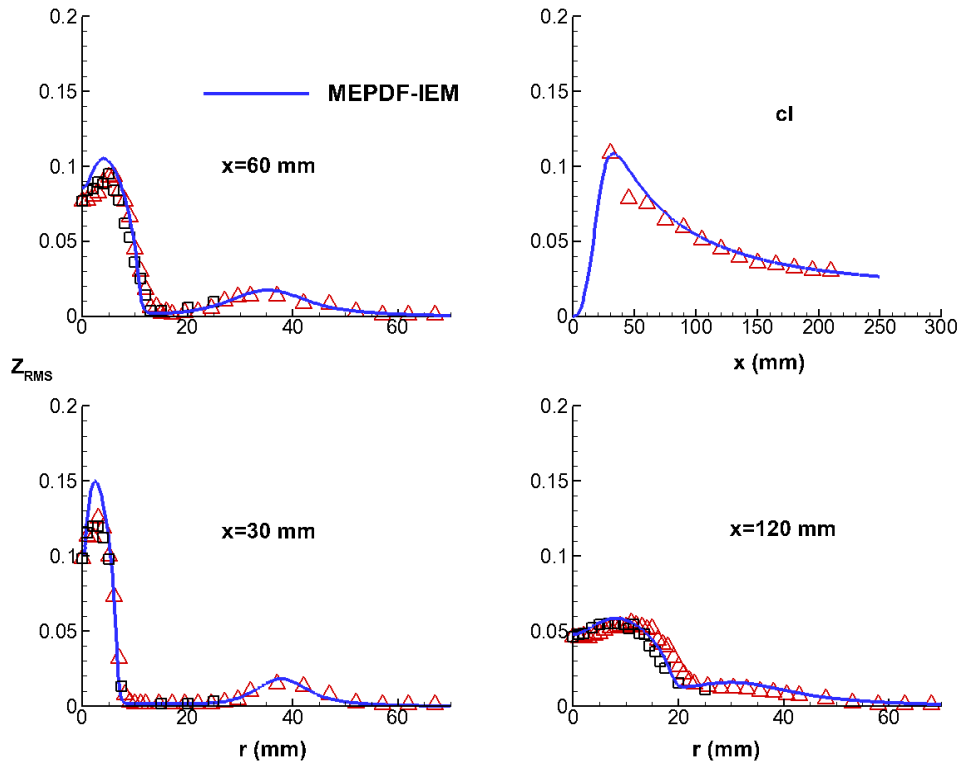

Figure 11: Radial and centerline RMS profiles of mixture fraction obtained for HM1 flames with $C_\varphi$=2. Symbols (*triangle* 0 ≤ r ≤ 70, *square* -25 ≤ r ≤ 0) are measurements and lines are predictions.

4.3 Results and discussion

The grid independence study in this case is carried out using two grids where numerical simulations are performed for HM1 flame with 3% oxygen in the coflow at Reynolds number, Re=10000. The coarse grid consists of 400x130 (axial x radial) cells whereas the finer grid has 800x240 cells. Figure 8 shows the results obtained using the MEPDF combustion model along with modified SKE turbulence model and DRM 19 [21] chemical mechanism using both the coarse and finer grids. As it is observed in the figure that the center line mean mixture fraction, obtained using above mentioned grids, are in good agreement with each other and measurements as well. Since the simulations using both the grids yield the same results, the coarse grid (400x130 cells) is chosen for the rest of the calculations.

We can evaluate the predictive capabilities of the MEPDF model in the downstream region along the burner axis through the center line profiles depicted in Figure 9. As observed from the figure, the model accurately predicts the mean mixture fraction as well as methane mass fraction profiles whereas the temperature profiles are under-predicted along the center line in the downstream region and the $O_2$ profiles are significantly over-predicted.

Figure 10 depicts the radial profiles of mean mixture fraction and mean temperature obtained for HM1 (3% $O_2$) flame. As observed from the radial profiles, the MEPDF model accurately captures the evolution of mean mixture fraction. The model produces agreeable temperature profiles till x=60mm, with slight under-estimation in the inner shear layer, after which they are significantly under-predicted in the downstream region. The peak temperature obtained in this case is approximately 1250K which is 10% lower than the experimental observations [1] and it is observed at radial location at x=30mm. While comparing the same with the published literature by Mardani et al. [10], EDC predictions also exhibit the similar peak temperature at x=30 mm.

Since this particular flame (HM1) poses extremely challenging case for modelling due to ultra-low $O_2$ content in the coflow, the under-predictions in the temperature profiles, obtained here, can be attributed to the handling of the correction terms in eqn. (4), which tries to de-stabilize the solution. When the values of scalars in different environments approach each other, the correction terms become unbounded and try to de-stabilize the solution. To reduce the departure of scalars in different environments and stabilize the solution, these correction terms are bounded in physical space by the mixing model. Therefore inaccuracies in the mixing model can produce discrepancies between experimental measurements and numerical predictions.

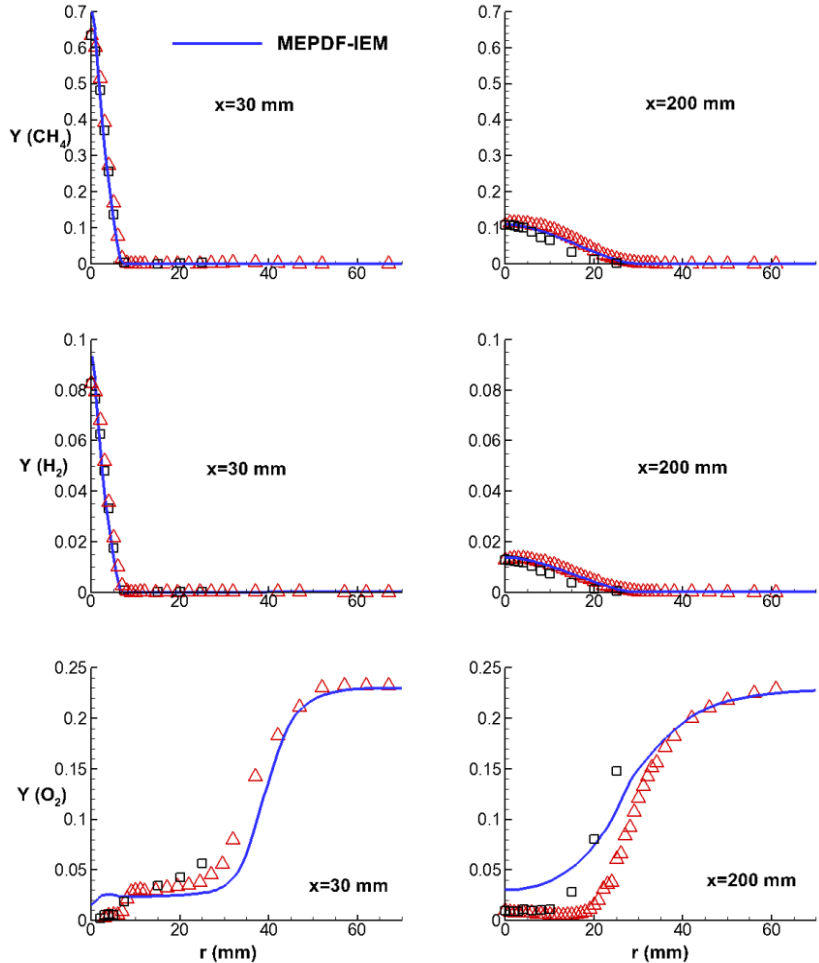

Figure 12: Radial profiles of $CH_4$, $H_2$, and $O_2$ mass fraction for HM1 flames with $C_\varphi=2$. Symbols (*triangle* $0 \leq r \leq 70$, *square* $-25 \leq r \leq 0$) are measurements and lines are predictions.

A better way to understand the predictive accuracy of the mixing model is to analyze the RMS profiles of mean mixture fraction. The mixture fraction here is the probability of the first environment and evolves due to turbulence only. While, the mean species mass fraction and the temperature are probability weighted averages from different environments. So, the error in the prediction of scalars can be due to the errors associated with the combustion modeling approach. In the present case, the sources of error are the micro mixing, which is modeled by IEM model. While the other source of error is the calculation of the correction sources, which is one of the modeling parameters of the MEPDF approach. The effect of inaccuracies in the mixture fraction due to the incorrect species and temperature prediction comes only through the density. The effect of inaccuracies of species and temperature on mean density calculations get diluted, hence the mixture fraction field is still reasonably good despite a larger inaccuracies in the scalar fields. In the current formulation, we are not solving the probability equation for all the environments. Therefore, we cannot get the variance of mixture fraction from the probability fields of different environments. However, from a given mean mixture fraction field, we can easily obtain the variances of mixture fraction by solving an extra scalar transport equation having convection, diffusion terms and a source term which is function of the gradient of mean mixture fraction. After solving this transport equation, we obtain the following mixture fraction variance field as shown in the figure (Fig. 11) below. We can see from the figure that the mixture fraction RMS is over predicted in the shear layer while away from it, the predictions are in excellent match. The over prediction of the mixture fraction in the shear layer is a solely due to the micro mixing model and has been in accordance with many published literature with IEM and Lagrangian PDF transport model.

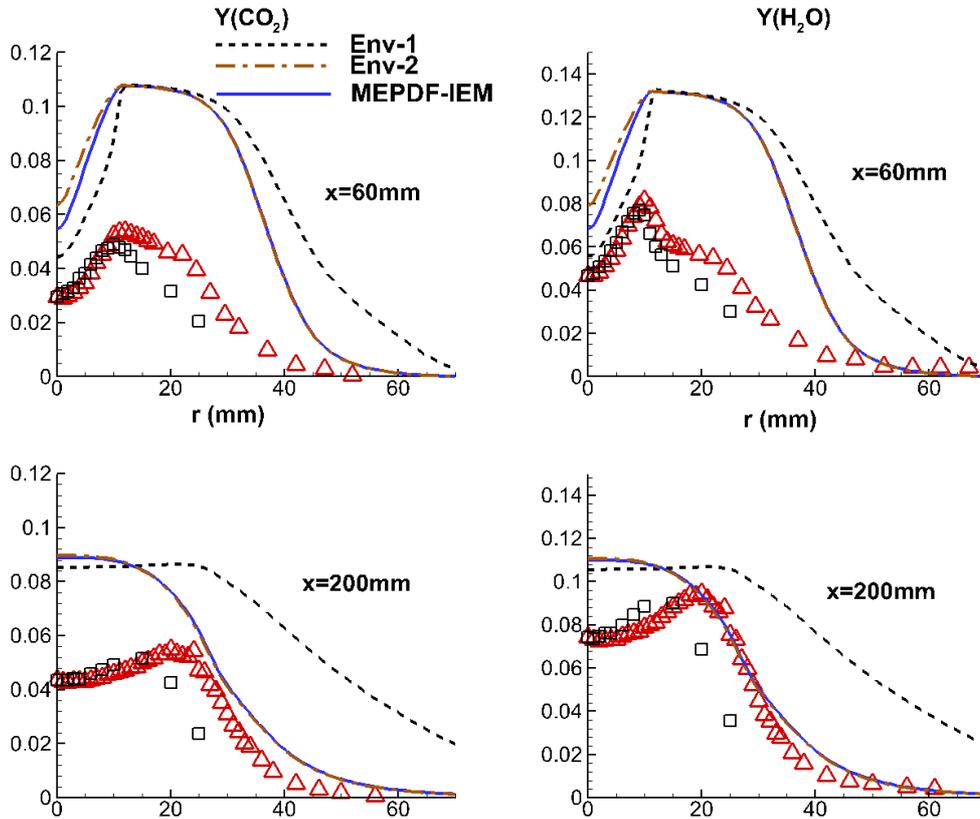

Figure 13: Radial profiles of $CO_2$ and $H_2O$ mass fraction for HM1 flames with $C_\varphi=2$. Symbols (*triangle* $0 \leq r \leq 70$, *square* $-25 \leq r \leq 0$) are measurements and lines are predictions.

The radial profiles of major species mean mass fraction ($CH_4$, $H_2$, and $O_2$) are depicted in Fig. 12. The $CH_4$ profiles are slightly over-predicted in the inner shear layer near the jet exit after which they are correctly predicted in the downstream region. A similar trend is observed for the $H_2$ profiles. On the other hand, the $O_2$ profiles are slightly over-predicted in the inner shear layer but are accurately captured in the outer shear layer. These discrepancies are observed due to error in predicting molecular mixing which can be associated with the inaccuracies of the IEM mixing model. A more accurate mixing model like Euclidean Minimum Spanning Tree (EMST) [20] can produce more accurate results compared to the IEM mixing model. But, the MEPDF model involves solution of transport equation in an Eulerian frame, while the EMST [20] model is a particle based model and hence cannot be used with MEPDF.

In the present MEPDF formulation, the energy and the species transport equations are solved independently for two environments. At the inlet boundary, environment-1 represents pure fuel whereas environment-2 represents pure oxidizer. The coupling between these environments takes place through the micro-mixing and correction terms. Figure 13 depicts the evolution of $CO_2$ and $H_2O$ species mass fraction in both the environments. The reaction, in any environment, will initialize when there is a sufficient amount of fuel and oxidizer present in the environment which will be followed by a chemical delay. The evolution of species in both environments takes place due to mixing and reaction in each environment as the species and energy diffuse across the environments due to mixing. Therefore, errors and inaccuracies of the mixing model will influence the results in both the environments significantly. As observed from the figure, both $CO_2$ and $H_2O$ species mass fractions are correctly captured by the model in environment-1 near the fuel-jet exit as environment-1 has a stronger effect on physical processes in the fuel-rich side on composition space. Whereas in environment-2, both the species mass fraction profiles are over-predicted as environment-2 dominantly influences the physical processes in the fuel-lean side. These two effects result in different flame evolution characteristics of each environment on the mixture fraction space.

The effect of differential diffusion has already been reported in the literature since the central fuel jet contains 11% $H_2$ by mass. However, it is always a challenging task to model flames in MILD regime due to slower reaction rate. Therefore, it is noteworthy to study the effects of reaction kinetics and differential diffusion as discussed in the next subsection.

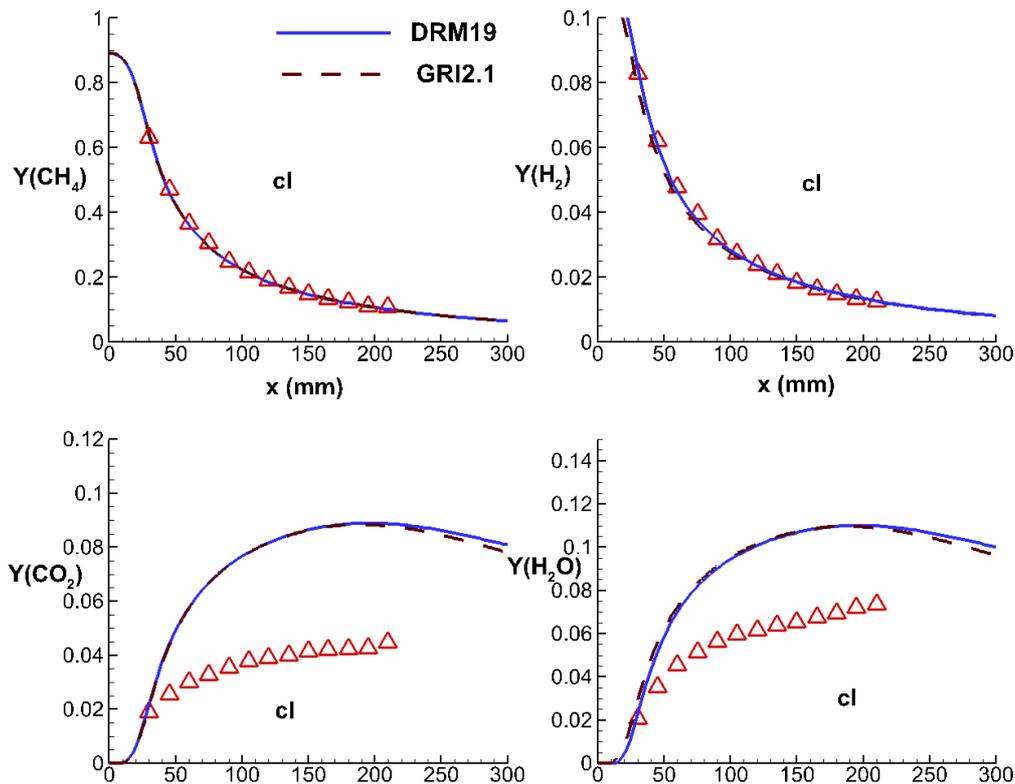

Figure 14: Centerline profiles of $CH_4$, $H_2$, $CO_2$, and $H_2O$ mass fraction for HM1 flames with $C_\varphi=2$. Symbols are measurements and lines are predictions.

4.3.1 Effects of chemical kinetics, multi-component diffusion and $C_\varphi$

The reduced chemical mechanisms have shorter ignition delays than that of detailed chemical mechanisms and, hence, they are more resilient to flame extinction. But in MILD combustion regime, it is desirable to have ignition delays so as to capture the effects of slower reaction rates. Thus, it is important to investigate the effects of detailed chemical mechanisms on the predictive capabilities of the model. The simulations for HM1 flame (3% $O_2$) are repeated with GRI 2.1 [40] detailed mechanism. Figure 14 shows the center line profiles of $CH_4$, $H_2$, $CO_2$, and $H_2O$ mass fraction obtained using both DRM 19 [21] and GRI 2.1 [40] mechanisms. As expected, the mass fraction of $CH_4$ and $H_2$ predictions do not exhibit any substantial differences using these two mechanisms, whereas $CO_2$ and $H_2O$ profiles show a little variation in the downstream region while significantly over-predicting the profiles. Results obtained using both the mechanisms have good agreement with each other in the fuel rich combustion zone confirming that reduced mechanisms can be used to model these flames without sacrificing much of the accuracy and the discrepancies observed are not due to chemical mechanisms.

The effects of multi-component diffusion have also been reported for HM1 flame [1]. This can have impact on the spatial transport of the species as well as in the joint scalar distribution due to micro mixing term. In the current work, only IEM model has been used, which ignores the multi-component diffusion. However, the diffusivity is calculated using the molecular kinetic theory to see the impact on the flow field. Therefore, the effect of multi-component diffusion is considered only in the spatial transport within an environment due to different diffusivities of the mixture components. Figure 15 depicts the radial as well as center line profiles of mean temperature and $CH_4$ mean mixture fraction for HM1 flame. Even though the species profiles doesn't show any substantial difference between the constant diffusivity and kinetic theory results, the temperature profiles show a temperature peak conforming the experimental profiles to slight under-prediction in the inner shear layer.

Similar to the DJHC flames, here also, a parametric study is performed using different values of micro-mixing constant coefficient, $C_\varphi$. Figure 16 reports the center line profiles of mean temperature along with $CH_4$ and $H_2$ mean species mass fraction profiles with different values of mixing constant coefficient, $C_\varphi$. It should be noted from the figure that for this case as well, changing the value of micro mixing constant coefficient leads to no substantial changes in the mean species profiles. The effect of radiative heat losses on the predictions is also studied by using the P1 radiation model and no substantial differences are observed between the results without any radiation model and the ones with P1 radiation model.

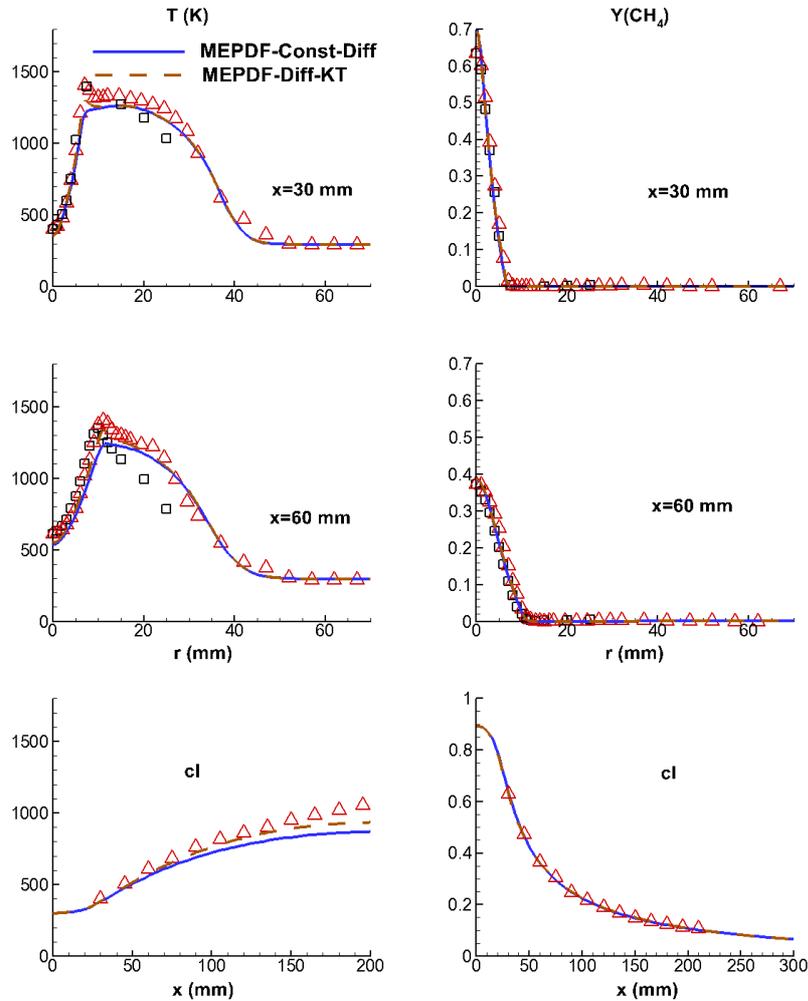

Figure 15: Radial and centerline profiles of mean temperature and $CH_4$ mass fraction for HM1 flames with $C_\varphi$=2. Symbols (*triangle* 0 ≤ r ≤ 70, *square* -25 ≤ r ≤ 0) are measurements and lines are predictions.

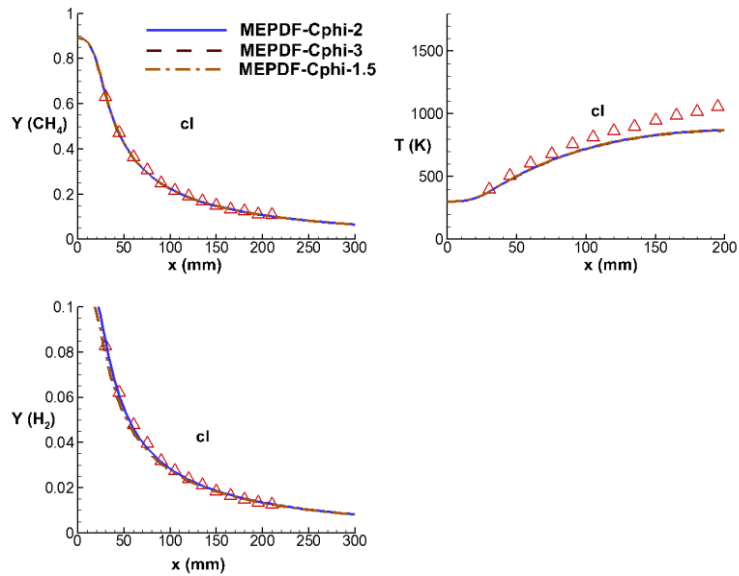

Figure 16: Centerline profiles of mean temperature and $CH_4$, $H_2$ mass fraction for HM1 with different $C_\varphi$. Symbols are experimental measurements and lines are predictions.

## 5. CONCLUSIONS

Two different burners imitating MILD combustion have been numerically investigated using MEPDF approach and reported in the current work. Initially, DJHC flames are simulated using two environments along with constant mixing co-efficient ($C_\varphi=2$). For Re=4100, the predictions are found to be in good agreement with the experimental database with temperature being an exception which is over-predicted significantly in the downstream region. Increasing the Reynolds number to Re=8800 increases the entrainment of the hot coflow into the fuel jet, thereby reducing the flame lift-off height which is nicely captured by the MEPDF model. The discrepancies in the predictions for higher Re case is increased and that is primarily due to the inaccuracies in the turbulence model and inappropriate response of the MEPDF model through the micro-mixing term. No substantial changes are observed while studying the effects of $C_\varphi$ on the flame characteristics. For higher O₂% (DJHC-X flame) case, the predictions are also in well agreement.

In the later part of the paper, the HM1 flame from the Adelaide burner are simulated and mean temperature along with mean mass fraction profiles of $CH_4$, $CO_2$, $H_2$, $H_2O$, and $O_2$ are reported at different axial and radial locations. The predictions using MEPDF model are found to be in excellent agreement apart from few discrepancies which can be attributed to the errors in turbulence model, sensitivity of the predictions to the handling of correction terms ($b_n$), and mixing model. No substantial differences are observed while studying the effects of multi-component diffusion on the flame characteristics; whereas we observe little discrepancies in the results while comparing the results from DRM19 and GRI 2.1 chemical mechanisms. Moreover, the variation in mixing model constant ($C_\varphi$) as well as radiation effect do not produce any substantial differences in the mean profiles. Overall, based on the results obtained it can be concluded that MEPDF model has the potential to be used as an alternative method to the Lagrangian PDF models for pure diffusion flames, while needs improvement for MILD flames.

## ACKNOWLEDGEMENTS

Simulations are carried out on the computers provided by IITK (www.iitk.ac.in/cc) and the manuscript preparation as well as data analysis is carried out using the resources available at IITK. This support is gratefully acknowledged. The authors would also like to thank Prof. Bassam Dally for providing the experimental database for the Adelaide burner.